\documentclass[aps,prd,twocolumn,10pt,altaffilletter,superscriptaddress,showpacs,nofootinbib,preprintnumbers]{revtex4-1}
\pdfoutput=1

\usepackage{graphicx,color,amsfonts,amssymb,lipsum,footnote,bm,keyval,hyperref,tikz,amsmath}
\hypersetup{pdfpagemode=UseNone}

\newcommand{\be}{\begin{equation}} 
\newcommand{\ee}{\end{equation}} 
\newcommand{\bea}{\begin{eqnarray}} 
\newcommand{\eea}{\end{eqnarray}} 
\newcommand{\lsq}{\left[} 
\newcommand{\rsq}{\right]} 
\newcommand{\lpa}{\left(} 
\newcommand{\rpa}{\right)} 
\newcommand{\nn}{\nonumber} 
\newcommand{\labsec}[1]{\label{sec:{#1}}} 
\newcommand{\secn}[1]{(\ref{sec:{#1}})} 
\newcommand{\labeq}[1]{\label{eq:{#1}}} 
\newcommand{\eqn}[1]{(\ref{eq:{#1}})} 
\newcommand{\Tsd}{\mbox{\fontsize{12pt}{0cm}\selectfont{$\mathrm{\mbox{$f$}}$}}} 
\newcommand{\qln}{{\rm{ln}}_\rmi{q}\ } 
\newcommand{\Rea}[1]{\mathrm{Re}\left(#1\right)} 
\newcommand{\dq}{{\delta{q}}} 
\newcommand{\mintedim}[2]{{\int\kern-0.50em\mbox{{\small$\mathop{\frac{\mbox{{\small${\rm d^{#2}}\vect{#1}$}}}{\mbox{{\small$(2\pi)^{#2}$}}}}$}}\ }} 
\newcommand{\inteonedim}[1]{{\int_0^\infty\kern-1em\mbox{{\small${\rm d}{#1}$}}}} 
\newcommand{\intecontour}[2]{{\int_{{#1}-i\infty}^{{#1}+i\infty}{\kern-2.25em d{#2}/(2i\pi)}}} 
\newcommand{\rmi}[1]{{\mbox{\scriptsize{#1}}}} 
\newcommand{\vect}[1]{\bm{#1}} 

\begin{document}

\title{Analytic results for the Tsallis thermodynamic variables}

\author{Trambak~Bhattacharyya}
\email{trambak.bhattacharyya@uct.ac.za}
\affiliation{UCT-CERN Research Centre, University of Cape Town, Rondebosch 7701, South Africa}
\affiliation{Department of Physics, University of Cape Town, Rondebosch 7701, South Africa}

\author{Jean~Cleymans}
\email{jean.cleymans@uct.ac.za}
\affiliation{UCT-CERN Research Centre, University of Cape Town, Rondebosch 7701, South Africa}
\affiliation{Department of Physics, University of Cape Town, Rondebosch 7701, South Africa}

\author{Sylvain~Mogliacci}
\email{sylvain.mogliacci@gmail.com}
\affiliation{UCT-CERN Research Centre, University of Cape Town, Rondebosch 7701, South Africa}

\everymath{\displaystyle}

\begin{abstract}
We analytically investigate the thermodynamic variables of a hot and dense system, in the framework of the Tsallis non-extensive classical statistics. After a brief review, we start by recalling the corresponding massless limits for all the thermodynamic variables. We then present the detail of calculation for the exact massive result regarding the pressure -- valid for all values of the $q$-parameter -- as well as the Tsallis $T$-, $\mu$- and $m$- parameters, the former characterizing the non-extensivity of the system. The results for other thermodynamic variables, in the massive case, readily follow from appropriate differentiations of the pressure, for which we provide the necessary formulas.
For the convenience of the reader, we tabulate all of our results. A special emphasis is put on the method used in order to perform these computations, which happens to reduce cumbersome momentum integrals into simpler ones. Numerical consistency between our analytic results and the corresponding usual numerical integrals are found to be perfectly consistent.
Finally, it should be noted that our findings substantially simplify calculations within the Tsallis framework. The latter being extensively used in various different fields of science as for example, but not limited to, high-energy nucleus collisions, we hope to enlighten a number of possible applications.
\end{abstract}

\maketitle

\section{Introduction}

Phenomena best described by power law distributions are present in various branches of physics~\cite{NatureTirnakliBorges,GranularCombeEtAll,ExpeNonHEPConfirmationDouglasEtAll}, and the Tsallis power law distribution~\cite{TsallisOriginal} is able to describe a number of those~\cite{TheorPredicConfirmedLutz}.

One of the interesting features of this distribution is that it bridges the power law type of distributions to the exponential ones. As can be seen from Sec.~\secn{TsallisReview}, the Tsallis distribution itself is a power law for values of the $q$-parameter greater than one, while it reproduces the usual
Boltzmann-Gibbs distribution in the limit $q\rightarrow 1$.

This framework is then extensively used in order to describe transverse momentum distributions in high energy collisions. The PHENIX and STAR collaborations~\cite{STAR,PHENIX1,PHENIX2} at the Relativistic Heavy Ion Collider (RHIC) in BNL and the ALICE, ATLAS and CMS collaborations~\cite{ALICE_charged,ALICE_piplus,CMS1,CMS2,ATLAS,ALICE_PbPb} at the Large Hadron Collider (LHC) in CERN, have indeed pioneered the use of this distribution in the field of experimental particle physics. Successful descriptions of the experimental transverse momentum distribution, the longitudinal momentum fraction distribution as well as the rapidity distribution of hadrons, off the $e^+e^-$ and $p-p$ collisions have been obtained in Refs.~\cite{E+E-,R1,R2,R3,Ijmpa,Plbwilk,Marques}.

The distribution is essentially described by two parameters: The first one, namely $q$, allows for interpolating between a power like behavior and an exponential one, while the second is the Tsallis temperature $T$. It is also possible to supplement the parameters with a mass $m$ and a set of chemical potentials. In the present work we will restrict ourselves to one species, the generalization to a finite number of species is straightforward. Interpreting the $q$ and $T$ parameters in terms of the parameters of the usual Boltzmann-Gibbs distribution -- a question of much conceptual interest -- has been investigated and we point out Ref.~\cite{WilkWlodarczyk} for more information on the matter.

Before proceeding further, we would like to draw the attention of the reader on the long standing latent controversy about potential conceptual problems within this framework. The predictive power of the framework, for example, did receive some attention in Refs.~\cite{Dauxois,RapisardaPluchinoToDauxois,TsallisToDauxois}. This question had been raised upon the release of an interesting paper~\cite{HilhorstSchehr} where it has been explicitly shown, though in two specific examples only, that some expected to be power like distributions turned out to be actually exponential ones. As a matter of fact, the present work is intentionally void of any such investigations, as we do not aim at these questions at all. Instead, we take this framework as a starting point and use a rigorous method in order to access analytic results regarding the underlying thermodynamics. As it is presently widely used, we hope that our work will be able to help in a number of related applications and investigations.

As previously mentioned, the Tsallis thermodynamics defined for $q\geq 1$ reduces to the Boltzmann-Gibbs one in the limit $q\rightarrow 1$. And just like for the Boltzmann statistics, one can describe Tsallis thermodynamic variables such as the number density, the energy density, the pressure and the entropy density, for values of $q$ above one. We refer to Sec.~\secn{TsallisReview} for a more detailed introduction on this topic. As is well known, analytical expressions for the Boltzmann thermodynamic variables can be represented in terms of modified Bessel functions. There have been attempts to obtain their counterparts in the Tsallis framework, in the massive case, by Taylor expanding and further truncating the distribution~\cite{WilkOsada,TsallisIndore,Alberico} prior to integrate over the momentum. Up to recently, the state-of-the-art was an expansion up to and including the order ${\cal O}(q-1)^2$. However, such an early truncation of the series automatically affects the convergence properties of the results, which makes them ultimately restricted by ratios typically depending on values of the energy, the temperature and $q$~\cite{TsallisIndore}. The latter being usually obtained by fitting actual experimental data, there is no control on the range of values for $q$. Unfortunately, the typical $q$ fitted values are most often beyond reach in respect to the convergence of the truncated expansion. This is one of the main reasons for analytically investigating the full massive results, valid for all $q$.

In this article, we then give explicit analytical representations for the Tsallis pressure in the massive case (from which one can obtain any thermodynamic variables by simple differentiations), making use of the well known Mellin-Barnes (MB) contour integral representation (see Ref.~\cite{Smirnov} and Refs.~\cite{BoosDavydychev,DavydychevTausk} for reviews and references on this topic), applied to the Tsallis distribution prior to integrate over the momentum. In this way, the typically cumbersome momentum integral is drastically simplified by means of mapping to the complex plane. It is then rather simple to perform this integration which, upon a careful choice in wrapping the contour onto one side of the real axis, finally leads to a series of residues involving the Gamma function. These type of series usually admit analytic closed forms in terms of hypergeometric functions. For the sake of cross-checking, the numerical integration of these variables using for example Mathematica, matches perfectly the numerical evaluation of the analytic formulas. As a consequence, we particularly hope that our exact results will replace lengthy and time consuming sub-routines among usual fitting procedures or so, and regarding a number of phenomenological uses especially (but not only) relevant to the LHC and RHIC.

This paper is organized as follows. We start with a brief review of the definitions for the Tsallis thermodynamic variables in the next section~\secn{TsallisReview}, based on a form of the Tsallis distribution described in detail in Refs.~\cite{Worku1,Worku2,Worku3,Azmi1}. In Sec.~\secn{ReadyToUseFormulas}, we tabulate all of our forthcoming results, before to explicitly calculate the Tsallis thermodynamic variables in the massless limit, in Sec.~\secn{TsallisMassless}. We then generalize the calculation to the massive case and give the full detail of the computation for the corresponding Tsallis pressure, in Sec.~\secn{TsallisMassive}. Finally, in Sec.~\secn{Conclu}, we present a summary of our results.

\section{Review of the Tsallis Thermodynamics}
\labsec{TsallisReview}

The Tsallis thermodynamic quantities, for a system of massive particles, can be written as integrals over certain combinations of the Tsallis distribution $f$, the modulus of the momentum $p\equiv |\vect{p}|$ and the energy $E_\rmi{p}\equiv\sqrt{p^2+m^2}$. The distribution is defined for any $q\geq 1$ by
\be
\Tsd \equiv \left[1+(q-1)\ \frac{E_\rmi{p}(m)-\mu}{T}\right]^{-\frac{1}{q-1}},
\labeq{tsallis}
\ee
which, aside from $q$ and $T$ contains also the chemical potential $\mu$ as a parameter. It is now obvious that the above reduces to the Boltzmann-Gibbs exponential distribution, in the limit $q\rightarrow 1$.

It can be shown (see Ref.~\cite{Worku2} for more details) that the total entropy $S$, the total particle number $N$, the total energy $E$, and the pressure $P$ are given by
\bea
S = s V\,&\equiv&\,-gV \mintedim{p}{3}\bigg[\ \Tsd^q \qln\Tsd-\Tsd\ \bigg] \labeq{entropy},\\
N = n V\,&\equiv&\,gV \mintedim{p}{3}\Tsd^q \labeq{Number},\\
E = \epsilon V\,&\equiv&\,gV \mintedim{p}{3}E_\rmi{p}\ \Tsd^q \labeq{epsilon},\\
P\,&\equiv&\,g \mintedim{p}{3}\frac{p^2}{3~E_\rmi{p}}\ \Tsd^q \labeq{pressure},
\eea
where we note that the lower case letters stand for the corresponding densities, $V$ being the volume and $g$ the degeneracy factor. Let us, from now on, use the variable $\dq\equiv q-1$ instead of $q$, as it will turn out to simplify most of our expressions.

We notice that the $\qln$ function, present in Eq.~\eqn{entropy}, is a specific $q$-logarithm defined for $\dq\geq 0$ by
\be
\qln x\equiv\frac{1-x^{-\dq}}{\dq}.
\ee
It interpolates between polynomial functions and reduces to the natural logarithm, when $\dq\rightarrow 0$. We further notice that the $q$-logarithm and the distribution must satisfy the relation below, for any $\dq\geq 0$, that is both functions must be inverse of each other, like their analogs, and
\be
\qln\Tsd\equiv\frac{1-\left(\left[1+\dq\ X\right]^{\frac{1}{-\dq}}\right)^{-\dq}}{\dq} = -X\ ,
\labeq{relaqlogqexp}
\ee
where in practice $X\equiv(E_\rmi{p}-\mu)/T$.

We see that the above obviously holds for $\dq=0$, as the $q$-functions become respectively the natural logarithm and the exponential function. The latter being the argument of the former, the natural logarithm does not lay onto the branch cuts as the exponential is always positive regardless of the sign of $X$. However, when $\dq>0$, the nature of the involved functions changes drastically. Thus, we see that for Eq.~\eqn{relaqlogqexp} to hold, we need $1+\dq\ X>0$ so that both $q$-power functions are single valued for non-integer values of the respective powers. When $\dq>0$, this requirement amounts to the condition that either $E_\rmi{p}\geq\mu$ or $m\geq\mu$, or else $m<\mu$ and $\dq<T/(\mu-E_\rmi{p})$ if $E_\rmi{p}<\mu$.

Therefore, we see that in order to keep the $q$-logarithm statistically meaningful (as related to the entropy), we basically need to restrict ourselves from asymptotically dense systems or enforce a specific constraint on the $\dq$ variable. We shall then assume all the necessary conditions to be fulfilled, along the rest of the paper. Notice that the simpler case $m\geq\mu$ is by far more common in most of the practical situations. We will then focus, from now on, on that situation unless probing the massless limit. However, we point out that the analysis of our computations and results, especially regarding the convergence regions for $\dq$, relevant to the case $m<\mu$ can also be readily obtained.

Now, given the definitions~\eqn{entropy},~\eqn{Number},~\eqn{epsilon}, and~\eqn{pressure}, it is very easy to show, using integration by parts, that these integrals correspond to the physical quantities, and hence must obey the fundamental equation of thermodynamics
\be
\epsilon + P = T s + \mu n,
\labeq{FundThermoEq}
\ee
only if $\dq$ is constrained to be strictly smaller than $1/3$. We will come back later to this constraint on $\dq$, as it will explicitly appear when working out the analytic structures of the corresponding integrals in the massless case.

Therefore, from now on, we shall keep in mind that the consistency of the framework requires that we have $0\leq \dq <1/3$ together with the above constraints on the parameters, namely the mass and the chemical potential which we chose to be $m\geq \mu$. We point out that the latter is consistent with most of the physical situations that are to be encountered when applying the present results.

One can also show~\cite{Worku2} that the use of $\Tsd^q$ instead of $\Tsd$ in order to define the thermodynamic variables leads to the usual thermodynamic consistency conditions such as
\be
T=\left.\frac{\partial \epsilon}{\partial s}\right|_n,\ \ \mu=\left.\frac{\partial \epsilon}{\partial
n}\right|_s,\ \ n=\left.\frac{\partial P}{\partial \mu}\right|_T,\ \ \ s=\left.\frac{\partial P}{\partial
T}\right|_\mu.
\ee
From the first of the above equations, it is obvious that the variable $T$ appearing in Eq.~\eqn{tsallis} is a thermodynamic temperature, hence more than just another parameter.

Using the Tsallis distribution, the particle spectrum can be written by means of more appropriate variables, {\it{e.g.}}, in the context of High-Energy Physics (HEP), as
\bea
\frac{dN}{dp_T dy} &=& \frac{gV}{(2\pi)^2}\ p_T\ m_T\ \mathrm{cosh}y \nn\\
&&\left(1+\dq\ \frac{m_T\ \mathrm{cosh}y-\mu}{T}\right)^{-\frac{1+\dq}{\dq}},
\labeq{tsallisptdist}
\eea
where $p_T\equiv\sqrt{p_x^2+p_y^2}$ and $m_T\equiv\sqrt{p_T^2+m^2}$ are the transverse momentum and the transverse mass respectively, $m$ being the bare mass.

Since the value of $\dq$ is usually quite close to 0 in HEP, the Tsallis distribution can be Taylor expanded~\cite{WilkOsada,TsallisIndore,Alberico} to yield analytical approximations of the Tsallis thermodynamic variables. However, as previously mentioned, the (too early) termination of the series implies a certain number of constraints on the energy, the temperature and the $\dq$ values. This is of course due to the fact that a Taylor expansion of such a function, around $\dq\approx 0$, effectively amounts to a Taylor expansion around $\dq(m_T\ \mathrm{cosh}y-\mu)/T\approx 0$ (see Ref.~\cite{TsallisIndore} for more details).

Another simplification is possible, in the massless limit, where the Tsallis thermodynamic variables have been found to be analytically computable (see the Appendix~A of Ref.~\cite{Ishihara} for a detailed computation), and that will be the subject of our discussion in Sec.~\secn{TsallisMassless}. But before to do so, let us summarize all of our results, for the convenience of the readers interested in direct applications.

\section{Ready to use formulas}
\labsec{ReadyToUseFormulas}

\subsection{Thermodynamic variables for $m=0$ and $\mu=0$}

Below, we list the final results for the main thermodynamic variables in the massless case, and with vanishing chemical potential. Those read
\bea
P&=&\frac{g\ T^4}{6\pi^2}\ \ \frac{1}{(1-\dq)\ (\frac{1}{2}-\dq)\
(\frac{1}{3}-\dq)}, \labeq{MasslessResZeroMuP}\\
\epsilon&=&\frac{g\ T^4}{2\pi^2}\ \ \frac{1}{(1-\dq)\ (\frac{1}{2}-\dq)\
(\frac{1}{3}-\dq)}, \labeq{MasslessResZeroMuE}\\
s&=&\frac{2 g\ T^3}{3\pi^2}\ \ \frac{1}{(1-\dq)\ (\frac{1}{2}-\dq)\
(\frac{1}{3}-\dq)}, \labeq{MasslessResZeroMuS}\\
n&=&\frac{g\ T^3}{2\pi^2}\ \ \frac{1}{(1-\dq)\ (\frac{1}{2}-\dq)}.\labeq{MasslessResZeroMuN}
\eea
All results are valid for $0\leq \dq <1/3$, as required from the consistency of the framework.

\subsection{Thermodynamic variables for $m=0$ and $\mu\neq 0$}

Below, we list the final results for the main thermodynamic variables in the massless case, but with a finite chemical potential. Those read
\bea
P&=&\frac{g\ T^4}{6\pi^2}\ \ \frac{\lpa 1-\dq\
\frac{\mu}{T}\rpa^{\frac{3\dq-1}{\dq}}}{(1-\dq)\ (\frac{1}{2}-\dq)\
(\frac{1}{3}-\dq)}, \labeq{MasslessResNonZeroMuP}\\
\epsilon&=&\frac{g\ T^4}{2\pi^2}\ \ \frac{\lpa 1-\dq\
\frac{\mu}{T}\rpa^{\frac{3\dq-1}{\dq}}}{(1-\dq)\ (\frac{1}{2}-\dq)\
(\frac{1}{3}-\dq)}, \labeq{MasslessResNonZeroMuE}\\
s&=&\frac{g\ T^3}{6\pi^2}\ \ \frac{\lpa 4-\frac{\mu}{T}-\dq\ \frac{\mu}{T}\rpa\
\lpa 1-\dq\ \frac{\mu}{T}\rpa^{\frac{2\dq-1}{\dq}}}{(1-\dq)\ (\frac{1}{2}-\dq)\
(\frac{1}{3}-\dq)}, \labeq{MasslessResNonZeroMuS}\\
n&=&\frac{g\ T^3}{2\pi^2}\ \ \frac{\lpa 1-\dq\
\frac{\mu}{T}\rpa^{\frac{2\dq-1}{\dq}}}{(1-\dq)\ (\frac{1}{2}-\dq)}.\labeq{MasslessResNonZeroMuN}
\eea
Again, all results are valid for $0\leq \dq <1/3$ as required from the consistency of the framework, and this time the additional constraint $\dq<T/\mu$ needs to be supplemented (which happens to guaranty the absence of branch cuts, given the $\dq$-power functions containing ratios of such a combination $\dq\ \mu/T$).

\subsection{Pressure for $m\neq 0$ and $\mu\neq 0$ in the upper $\dq$-region}

Below, we display the final result for the full pressure in the massive case, with a finite chemical potential, evaluated in the upper $\dq$-region
\begin{widetext}
\bea
P_\rmi{U}&=&\frac{g\ m^4}{16 \pi^{\frac{3}{2}}}\ \left(\frac{T}{\dq\ m}\right)^{\frac{1+\dq}{\dq}}\ \left[%
\frac{\Gamma\left(\frac{1-3\dq}{2\dq}\right)}{\Gamma\left(\frac{1+2\dq}{2\dq}\right)}\times\,%
_2F_1\left(\frac{1+\dq}{2 \dq},\frac{1-3\dq}{2 \dq},\frac{1}{2};\left(\frac{\dq\ \mu-T}{\dq\ m}\right)^2\right)\right. \labeq{MassiveResPU}\\
&&\ \ \ \ \ \ \ \ \ \ \ \ \ \ \ \ \ \ \ \ \ \ \ \ \ \ \ \ \ +2\left.\left(\frac{\dq\ \mu-T}{\dq\ m}\right)\times\frac{\Gamma\left(\frac{1-2\dq}{2\dq}\right)}{\Gamma\left(\frac{1+\dq}{2\dq}\right)}\times\,%
_2F_1\left(\frac{1+2\dq}{2 \dq},\frac{1-2\dq}{2\dq},\frac{3}{2};\left(\frac{\dq\ \mu-T}{\dq\ m}\right)^2\right)%
\right],\nn
\eea
\end{widetext}
which is valid for $0\leq \dq <1/3$ required from the consistency of the framework, and $\dq>T/(m+\mu)$ from the convergence of the Gauss hypergeometric functions.

\subsection{Pressure for $m\neq 0$ and $\mu\neq 0$ in the lower $\dq$-region}

Below, we display the final result for the full pressure in the massive case, with a finite chemical potential, evaluated in the lower $\dq$-region
\begin{widetext}
\bea
P_\rmi{L}&=&\frac{g\ m^4}{\pi^{\frac{3}{2}}}\left(\frac{T/2}{T-\mu~\dq}\right)^{\frac{1+\dq}{\dq}}\ \left[%
\frac{\dq^2\ (2-\dq)\ \Gamma\left(\frac{1}{\dq}\right)}{(1-3\dq)\ (1-2\dq)\ (1-\dq)\ \Gamma\left(\frac{2+\dq}{2 \dq}\right)}%
\right] \labeq{MassiveResPL}\\
&&\ \ \ \ \ \ \ \ \ \ \ \ \ \ \ \ \ \ \ \ \ \ \ \ \ \ \ \ \ \ \ \ \ \times\ %
_2F_1 \left( \frac{1+2\dq}{2\dq},\frac{1+\dq}{2\dq},\frac{2-\dq}{2\dq}, 1-\left(\frac{\dq~ m}{T- \mu~\dq}\right)^2\right)%
,\nn
\eea
\end{widetext}
which is valid for $0\leq \dq <1/3$ required from the consistency of the framework, and $\dq \leq T/(m+\mu)$ from the convergence of the Gauss hypergeometric function.

\section{Thermodynamic variables in the massless limit}
\labsec{TsallisMassless}

Let us define a more general integral, encompassing all variables in the massless limits present in the literature, and reproducing Eqs.~\eqn{entropy},~\eqn{Number},~\eqn{epsilon}, and~\eqn{pressure}, by
\be
\mathcal{I}(\alpha,\beta)\equiv g\ \mintedim{p}{3}\frac{p^{\beta-2}}{\Big[1+\dq\
\frac{p-\mu}{T}\Big]^{\frac{\alpha}{\dq}}},
\labeq{intgeneralmassless}
\ee
where here, $\alpha$ and $\beta$ are nothing but just handy variables, the former being set to either $1+\dq$ or $1$ at the end, in order to recover the thermodynamic variables.

The above integral is built to converge, in three dimensions, upon some constraints on the various parameters. Those encompass the fact that the integrand shall only assume real values, and the usual infrared and ultraviolet convergence conditions. The conditions for the massless case turn to be
\bea
1+\Rea{\beta}>0\ &&,\ \ \Rea{\alpha}>0\ \labeq{SetConstrainsZeroM1},\\
T>\dq\ \mu\ &&,\ \ \dq<\frac{\Rea{\alpha}}{1+\Rea{\beta}}
\labeq{SetConstrainsZeroM2},
\eea
for which we see that the first two are trivially accomplished, given the actual relevant set of integrals we wish to compute. The last two, on the other hand, are not trivial at all. Given the usual $\alpha$ and $\beta$ values we are interested in, {\it{e.g.}}, some combinations of $\alpha=1+\dq,\ 1$ and $\beta=2,\ 3$, we see that $\dq$ must indeed be bounded at least by $\dq<1/3$, as we previously mentioned. This being said, the remaining constraint on the $T$ and $\mu$ parameters is either not needed in the case $E_\rmi{p}\geq\mu$, or fulfilled if $m<\mu$ with $\dq<T/(\mu-E_\rmi{p})$ and $E_\rmi{p}<\mu$, only here for $m=0$ in both cases. Notice that this last constraint brings an overall limit for the chemical potential, which is $\mu<3 T$. Thus, we see again that this framework must be kept away from asymptotically dense systems.

We are now going to investigate the general integral that one can use to both express all the thermodynamic variables, and account for the different massless variables used in the literature. Before doing so, let us notice that $n=\mathcal{I}(\alpha=1+\dq,\beta=2)$ for the particle number density, as well as $\epsilon=\mathcal{I}(\alpha=1+\dq,\beta=3)$ for the energy density and $P=\mathcal{I}(\alpha=1+\dq,\beta=3)/3$ for the pressure.

\subsection{Computing the general massless integral}
\labsec{generalmassless}

After having performed the angular momentum integration, and upon redefining the variable $p$ such that $p^{\prime}\equiv 1+\dq\ (p-\mu)/T$, we can compute~\eqn{intgeneralmassless} using an integral representation for the so-called Beta function
\be
B(a,b)\equiv\frac{\Gamma(a)\
\Gamma(b)}{\Gamma(a+b)}=\int_{0}^{\infty}\kern-0.50em\mbox{{\small$\mathop{\mbox
{{\small${\rm d}u$}}}$}}\ u^{a-1}\ (1+u)^{-(a+b)},
\ee
where the last equality is only valid for strictly positive real parts of both parameters. Doing so, we obtain
\bea
\mathcal{I}(\alpha,\beta)&=&\ \frac{g\ T^{1+\beta}}{2\pi^2}\ \ %
\frac{\lpa 1-\dq\frac{\mu}{T}\rpa^{1+\beta-\frac{\alpha}{\dq}}}{\dq^{1+\beta}}
\labeq{resultintgeneralmassless}\ \ \ \ \nn\\
&&\ \ \ \ \ \times \frac{\Gamma(\frac{\alpha}{\dq}-1-\beta)\
\Gamma(1+\beta)}{\Gamma(\frac{\alpha}{\dq})},
\eea
with the set of constraints~\eqn{SetConstrainsZeroM1}
and~\eqn{SetConstrainsZeroM2}. We then notice our result agrees with the one previously derived in~\cite{Ishihara}.

\subsection{Number density}
\labsec{ndensitymassless}

In the massless limit, the number density can then be obtained from Eq.~\eqn{resultintgeneralmassless} by setting $\alpha=1+\dq$ and $\beta=2$. Simplifying the arguments of some of the Gamma functions, we then arrive at
\bea
n&=&\frac{g\ T^3}{2\pi^2}\ \ \frac{\lpa 1-\dq\
\frac{\mu}{T}\rpa^{\frac{2\dq-1}{\dq}}}{(1-\dq)\
(\frac{1}{2}-\dq)}.\labeq{ntsallismassless}
\eea

From the above expression, wee see that the number density is divergent at $\dq=1/2,~1$ which in the light of the previous discussion, implies that the condition of convergence for the corresponding integral is $0\leq \dq<1/2$. However, we recall that for $\dq$ above $1/3$, other thermodynamic quantities do not make sense anymore, as we are going to see in the next subsections. Therefore, the physically relevant range remains $0\leq \dq<1/3$, which is fairly enough to encompass values relevant to HEP.

\subsection{Energy density}
\labsec{edensitymassless}

Similarly, the massless energy density can be obtained from Eq.~\eqn{resultintgeneralmassless} by setting $\alpha=1+\dq$ and $\beta=3$. Using the same tricks as above, we then arrive at
\bea
\epsilon&=&\frac{g\ T^4}{2\pi^2}\ \ \frac{\lpa 1-\dq\
\frac{\mu}{T}\rpa^{\frac{3\dq-1}{\dq}}}{(1-\dq)\ (\frac{1}{2}-\dq)\
(\frac{1}{3}-\dq)}.\labeq{etsallismassless}
\eea
This time, from the above expression, wee see that the energy density diverges at $\dq=1/3,~1/2,~1$, which again shows that the condition of convergence for the corresponding integral is $0\leq \dq<1/3$, precisely the physically relevant range that we just mentioned.

\subsection{Pressure}
\labsec{pressuremassless}

In the massless limit, the pressure can be obtained from Eq.~\eqn{resultintgeneralmassless} by setting $\alpha=1+\dq$ and $\beta=3$, and dividing by $3$. Using the same methods as above, this gives us
\bea
P&=&\frac{g\ T^4}{6\pi^2}\ \ \frac{\lpa 1-\dq\
\frac{\mu}{T}\rpa^{\frac{3\dq-1}{\dq}}}{(1-\dq)\ (\frac{1}{2}-\dq)\
(\frac{1}{3}-\dq)},\labeq{ptsallismassless}
\eea
and we see that for a system of massless free classical particles following the Tsallis Statistics, we indeed have $P=\epsilon/3$. In this case also, divergences arise at $\dq=1/3,~1/2,~1$, and the range of convergence is the physically relevant one $0\leq \dq<1/3$.

\subsection{Entropy density}
\labsec{sdensitymassless}

Finally, we note from Eq.~\eqn{entropy} that the entropy density can also be straightforwardly obtained as
\bea
s&=&\frac{1+\dq}{\dq}\
\mathcal{I}(\alpha=1,\beta=2)-\frac{\mathcal{I}(\alpha=1+\dq,\beta=2)}{\dq}.\ \
\ \ \ 
\eea
By doing so, we get
\be
s=\frac{g\ T^3}{6\pi^2}\ \ \frac{\lpa 4-\frac{\mu}{T}-\dq\ \frac{\mu}{T}\rpa\
\lpa 1-\dq\ \frac{\mu}{T}\rpa^{\frac{2\dq-1}{\dq}}}{(1-\dq)\ (\frac{1}{2}-\dq)\
(\frac{1}{3}-\dq)}, \ \ \ \labeq{stsallismassless}
\ee
with again divergences at $\dq=1/3,~1/2,~1$, which are avoidable if $\dq$ is in the physically relevant range $0\leq \dq<1/3$. Notice that the above result matches the entropy obtained by plugging in all the results but the entropy, into the fundamental thermodynamic equation~\eqn{FundThermoEq}.

We finally notice that the above results satisfy the fundamental equation of thermodynamics, and that each variables is related to the pressure via the relevant derivative, as it should be.

\section{The pressure for systems with massive particles}
\labsec{TsallisMassive}

Unlike usual situations when applying MB techniques, we will not keep the number of spatial dimensions arbitrary as our integrals are defined to be convergent in the physically acceptable range $0\leq \dq<1/3$.

In the following, we shall always assume the above constraint, consistent with the Tsallis statistics, to be true (see Sec.~\secn{TsallisReview} for more details). We also chose, for the sake of argument, $m\geq\mu$ --- keeping in mind that situations with bigger chemical potential than the mass can easily be implemented as well.

We now turn toward the integral in Eq.~\eqn{pressure}, integrate over the angular part and performe the change of variable $p\rightarrow k\equiv p/m$. Doing so, the expression for the massive pressure can be rewritten as
\begin{widetext}
\be
P=\frac{g\ m^{4}}{6\pi^2}\ \inteonedim{k}\left[\frac{k^{4}}{\sqrt{1+k^2}}\times%
\frac{1}{\bigg[\left\{1-\dq\ \frac{\mu}{T}\right\}+\left\{\dq\ \frac{m}{T}\
\sqrt{1+k^2}\right\}\bigg]^{\frac{1+\dq}{\dq}}}\right].\labeq{kintegral}
\ee
\end{widetext}

We then recall the MB contour integral representation (again, see~\cite{BoosDavydychev},~\cite{DavydychevTausk}and~\cite{Smirnov} for more details)
\be
\frac{1}{(X+Y)^{\lambda}}=\intecontour{\epsilon}{z}\lsq\frac{\Gamma(-z)\
\Gamma(\lambda+z)}{\Gamma(\lambda)}\ \frac{Y^z}{X^{\lambda+z}}\rsq,
\ee
valid here for $\Rea{\lambda}>0$ and $\Rea{\epsilon}\in (-\mathrm{Re}(\lambda),0)$. Notice that in the present case, $\lambda$ has no imaginary part.

We can now apply the above formula to the $\dq$-dependent denominator in~\eqn{kintegral}, since $(1+\dq)/\dq>0$ is fulfilled given the previous assumptions. We do so with $X=\dq\ m/T\ \sqrt{1+k^2}$, $Y=1-\dq\ \mu/T$, and $\lambda=(1+\dq)/\dq$, change the order between the contour and the momentum integrals relying upon the convergence of the involved expressions, and finally obtain
\begin{widetext}
\be
P=\frac{g\ m^{4}\lpa \dq\ \frac{m}{T}\rpa^{-\frac{1+\dq}{\dq}}}{2\pi^2\
\Gamma\lpa\frac{1+\dq}{\dq}\rpa}%
\intecontour{\epsilon}{z}\lsq\lpa\frac{T-\dq\ \mu}{\dq\ m}\rpa^{z}\Gamma(-z)\
\Gamma\lpa z+\frac{1+\dq}{\dq}\rpa%
\inteonedim{k}\lsq\frac{k^{4}}{\lpa\sqrt{1+k^2}\rpa^{\frac{1+2\dq}{\dq}+z}}
\rsq\rsq.\labeq{ContourAndMomentumIntegrals}
\ee
\end{widetext}

We now want to perform the momentum integral in~\eqn{ContourAndMomentumIntegrals}. We see that this integral introduces another parameter, namely $z$, for which another constraint will be needed. Given the constraint which as previously seen naturally introduces an upper bound $\dq<1/3$, we see that for the above momentum integral to be convergent we only further need $\Rea{z}\geq 0$. This new constraint clearly indicates that for further performing the last $z$-integral by wrapping the contour onto one of the two sides of the real axis, only one side will be allowed if we are not to analytically continue prior to wrap the contour: The one for which $\Rea{z}\geq 0$. Before to wrap the contour then, we shall keep the parameters and especially $\dq$ arbitrary, in such a way that the momentum integral remains convergent (basically, $\dq$ is kept far from $\dq=0$). Then, when the contour is wrapped, we will be able to relax this arbitrary constraint.

In addition, unlike with usual MB representations, one of our parameter, namely $\dq$, is both present as a power and as a multiplicative factor that must control the convergence of the series of residues, when wrapping the contour to explicitly compute the $z$-integral. This last factual point complicates to quite some extent the use of this procedure, as it effectively introduces a non-trivial restricted range of validity for $\dq$, within the physically acceptable range $0\leq \dq<1/3$ --- if we are to obtain a closed form expression for the thermodynamic functions valid at least somewhere in the physical range.

Within usual MB procedures, as the dimension of the space $D$ is kept arbitrary, one can actually analytically continue the integrand of the contour integral as a function of $D$, prior to wrap the contour onto the side which was originally forbidden by the constraint on $z$, and obtain the corresponding closed form in the complementary part of the restricted range for $\dq$. Note that if we do so, we could access the originally forbidden $\dq$-region and obtain the corresponding analytic result, only at the cost not to obtain a closed form expression, as a matter of fact in the present case. Consequently, as we chose not to keep the dimension arbitrary, we cannot proceed in the usual manner. However, this does not mean that we will not be able to access the complementary part of the restricted range over $\dq$, and in the end obtain a set of closed forms for the massive pressure within the whole physical range for $\dq$. To do so, in the forthcoming subsection, we will have to analytically continue the final closed result for the pressure -- and not the integrand prior to wrap the contour -- to the complementary $\dq$-region. In this way, we will obtain a set of two formulas valid in two different ranges within $0\leq \dq<1/3$, both being complementary from each others.

Following the above procedure, and keeping in mind that we must have $\dq\geq 0$ such that the momentum integral does not diverge before wrapping the contour on the right positive part of the real $z$-axis, we obtain
\begin{widetext}
\be
P=\frac{3\ g\
m^{4}}{16\pi^{\frac{3}{2}}\lpa\dq\frac{m}{T}\rpa^{\frac{1+\dq}{\dq}}
\Gamma\lpa\frac{1+\dq}{\dq}\rpa}%
\intecontour{\epsilon}{z}\lsq\lpa\frac{T-\dq\ \mu}{\dq\
m}\rpa^{z}\times\frac{\Gamma(-z)\ \Gamma\lpa
z+\frac{1+\dq}{\dq}\rpa\Gamma\lpa\frac{z}{2}+\frac{1-3\dq}{2\dq}\rpa}{
\Gamma\lpa\frac{z}{2}+\frac{1+2\dq}{2\dq}\rpa}\rsq,\labeq{ContourIntegral1}
\ee
\end{widetext}
where we recall that we must close the contour onto the right side, keeping then $\Rea{z}\geq 0$. The need to close the contour onto the right side, leads to an additional condition for the power inside the integrand whose absolute value must be then smaller than one, if we are to obtain a convergent subsequent series over residues. This further constraint amounts to consider $\dq$ such that
\be
\dq>\frac{T}{m+\mu} \labeq{UpperRegion1}
\ee
given our choice $m\geq\mu$, and with the overall requirement that $\dq$ still belongs to the physical region $0\leq \dq<1/3$.

We will name~\eqn{UpperRegion1} the upper $\dq$-region investigated in the next subsection, its counter part being the lower $\dq$-region which we will further investigate in the following subsection~\secn{PressuremassiveLower}.

At last for now, we shall perform the change of variable $z\rightarrow 2 z$ and apply the so-called duplication formula to some of the Gamma functions, in order to simplify their arguments. Doing so, Eq.~\eqn{ContourIntegral1} then becomes
\begin{widetext}
\bea
P&=&\frac{3\ g\ \lpa\dq\frac{m}{2T}\rpa^{-\frac{1+\dq}{\dq}}}{32\pi^{\frac{5}{2}}\ m^{4}\Gamma\lpa\frac{1+\dq}{\dq}\rpa}\intecontour{\frac{\epsilon}{2}}{z}\lsq\lpa\frac{T-\dq\ \mu}{\dq\ m}\rpa^{2z}\times\frac{1}{\Gamma\lpa z+\frac{1}{2}+\frac{1+\dq}{2\dq}\rpa}\right.\times\\
&&\ \ \ \ \ \ \ \times\left.\Gamma\lpa -z\rpa\ \Gamma\lpa\frac{1}{2}-z\rpa\ \Gamma\lpa z+\frac{1+\dq}{2\dq}\rpa\ \Gamma\lpa \frac{1}{2}+z+\frac{1+\dq}{2\dq}\rpa\Gamma\lpa z-2+\frac{1+\dq}{2\dq}\rpa\rsq,\labeq{ContourIntegral2}
\eea
\end{widetext}
a representation which will be used for closing the contour to the right side for which $\Rea{z}\geq 0$, in the sub-region of $0\leq \dq<1/3$ for which~\eqn{UpperRegion1} is fulfilled, and of course given the previous assumptions on the mass and the chemical potential.

Notice that we should the encounter two distinct series of residues from the poles of $\Gamma(-z)$ and $\Gamma(1/2-z)$, respectively, when closing the contour onto the right side as will be detailed in the next subsection.

\subsection{Pressure in the upper q-region}
\labsec{PressuremassiveUpper}

Closing the contour onto the right side in~\eqn{ContourIntegral2} and further simplifying the integrand, the resulting series representation for the pressure turns to be
\begin{widetext}
\bea
P_{U}&=&\frac{g\ m^4\ \lpa\dq\ \frac{m}{2T}\rpa^{-\frac{1+\dq}{\dq}}}{32\
\pi^{\frac{5}{2}}\ \Gamma\lpa\frac{1+\dq}{\dq}\rpa}%
\sum_{k=0}^{\infty}\lsq\frac{(-1)^k}{k!}\ \lpa\frac{T-\dq\ \mu}{\dq\ m}\rpa^{2k}\Gamma\lpa\frac{1}{2}-k\rpa\ \Gamma\lpa k+\frac{1+\dq}{2\dq}\rpa\ \Gamma\lpa k-2+\frac{1+\dq}{2\dq}\rpa\rsq\nn\\
&+&\frac{g\ m^4\ \lpa 1-\dq\frac{\mu}{T}\rpa}{64\ \pi^{\frac{5}{2}}\ \Gamma\lpa\frac{1+\dq}{\dq}\rpa\lpa\dq\frac{m}{2T}\rpa^{\frac{1+3\dq}{2\dq}}}\times\nn\\
&&\ \ \ \ \ \ \ \ \ \ \ \ \times\sum_{k=0}^{\infty}\lsq\frac{(-1)^k}{k!}\ \lpa\frac{T-\dq\ \mu}{\dq\ m}\rpa^{2k}\Gamma\lpa -\frac{1}{2}-k\rpa\ \Gamma\lpa k+\frac{1}{2}+\frac{1+\dq}{2\dq}\rpa\ \Gamma\lpa k-\frac{3}{2}+\frac{1+\dq}{2\dq}\rpa\rsq,\labeq{SeriesP}
\eea
\end{widetext}
valid in the upper $\dq$-region~\eqn{UpperRegion1} of $0\leq \dq<1/3$, given the previous assumptions on $m$ and $\mu$.

Finally, the above series representation admits the following closed form, which we 'aesthetically improved'
\begin{widetext}
\bea
P_\rmi{U}&=&\frac{g\ m^4}{16 \pi^{\frac{3}{2}}}\ \left(\frac{T}{\dq\ m}\right)^{\frac{1+\dq}{\dq}}\ \left[%
\frac{\Gamma\left(\frac{1-3\dq}{2\dq}\right)}{\Gamma\left(\frac{1+2\dq}{2\dq}\right)}\times\,%
_2F_1\left(\frac{1+\dq}{2 \dq},\frac{1-3\dq}{2 \dq},\frac{1}{2};\left(\frac{\dq\ \mu-T}{\dq\ m}\right)^2\right)\right.\\
&&\ \ \ \ \ \ \ \ \ \ \ \ \ \ \ \ \ \ \ \ \ \ \ \ \ \ \ \ \ +2\left.\left(\frac{\dq\ \mu-T}{\dq\ m}\right)\times\frac{\Gamma\left(\frac{1-2\dq}{2\dq}\right)}{\Gamma\left(\frac{1+\dq}{2\dq}\right)}\times\,%
_2F_1\left(\frac{1+2\dq}{2 \dq},\frac{1-2\dq}{2\dq},\frac{3}{2};\left(\frac{\dq\ \mu-T}{\dq\ m}\right)^2\right)%
\right],\nn
\eea
\end{widetext}
in terms of the so-called Gauss hypergeometric function. This is the first closed formula for the pressure in the massive case, and it is valid in the upper $\dq$-region~\eqn{UpperRegion1} of $0\leq \dq<1/3$, in the present case given the previous assumptions on the mass and chemical potential. We notice that the above formula can be extended outside the physically relevant region, as long as it remains inside the region~\eqn{UpperRegion1}, and up to the few isolated poles such as $\dq=1/3$, $\dq=1/2$, and $\dq=1$. However, the resulting analytic result cannot be interpreted as the Tsallis pressure anymore, as the fundamental thermodynamic relation~\eqn{FundThermoEq} is then not satisfied.

\subsection{Pressure in the lower q-region}
\labsec{PressuremassiveLower}

Using the analytic continuation of the Hypergeometric functions from~\cite{bateman}, we get the pressure in the region $\dq \leq T/(m+\mu)$ which reads, after some more simplifications
\begin{widetext}
\bea
P_\rmi{L}&=&\frac{g\ m^4}{\pi^{\frac{3}{2}}}\left(\frac{T/2}{T-\mu~\dq}\right)^{\frac{1+\dq}{\dq}}\ \left[%
\frac{\dq^2\ (2-\dq)\ \Gamma\left(\frac{1}{\dq}\right)}{(1-3\dq)\ (1-2\dq)\ (1-\dq)\ \Gamma\left(\frac{2+\dq}{2 \dq}\right)}%
\right]\\
&&\ \ \ \ \ \ \ \ \ \ \ \ \ \ \ \ \ \ \ \ \ \ \ \ \ \ \ \ \ \ \ \ \ \times\ %
_2F_1 \left( \frac{1+2\dq}{2\dq},\frac{1+\dq}{2\dq},\frac{2-\dq}{2\dq}, 1-\left(\frac{\dq~ m}{T- \mu~\dq}\right)^2\right)%
,\nn
\eea
\end{widetext}
provided that $ \dq~ \mu< T$, which is automatically satisfied since in this region we have $(m+\mu) \dq \leq T$.

\section{Other Thermodynamic Variables}

Other thermodynamic variables like the number density ($n$), the entropy density ($s$) and the energy density ($\epsilon$) can be obtained using the following relations
\bea
n&=&\left.\frac{\partial P}{\partial \mu}\right|_{T},~
s=\left.\frac{\partial P}{\partial T}\right|_{\mu}, \nn\\
\epsilon&=&T\left.\frac{\partial P}{\partial
T}\right|_{\mu}+\mu \left.\frac{\partial P}{\partial \mu}\right|_{T} - P.
\eea
The derivatives of the hypergeometric functions (with respect to $T$ or $\mu$) appearing inside the analytical expression of pressure can be computed using the following chain rule (here for the temperature only)
\bea
\frac{\partial~ _2F_1 [a,b,c;f(T,\mu)]}{\partial T}=\frac{\partial~ _2F_1
[a,b,c;f(T,\mu)]}{\partial f}~\frac{\partial
f}{\partial T}, \nn\\
\eea
noticing the fact that
\bea
\frac{\partial~ _2F_1
[a,b,c;f]}{\partial f}=\frac{a~b}{c}\ ~_2F_1
[a+1,b+1,c+1;f]. \nn\\
\eea

\section{Summary}
\labsec{Conclu}

The thermodynamic variables which arise in Tsallis non-extensive thermodynamics have been investigated analytically. For the massless case with zero chemical potential, simple analytic expressions were obtained and presented in Eqs.~\eqn{MasslessResZeroMuP} to~\eqn{MasslessResZeroMuN}, as well as in Eqs.~\eqn{MasslessResNonZeroMuP} to~\eqn{MasslessResNonZeroMuN} when considering a non-zero chemical potential. Limits on the non-extensive variable $q$ have been presented, leading to a physically meaningful range of $0\leq q < 4/3$. The case of massive particles turns to be considerably more involved, and use was made of the Mellin-Barnes contour integral representation.

Explicit and detailed analytic results have been presented for the pressure in Eqs.~\eqn{MassiveResPU} and~\eqn{MassiveResPL}. The corresponding results for the number density, the energy density and the entropy density can be obtained in a straightforward manner by taking appropriate derivatives of the pressure. The results obtained reduce cumbersome momentum integrals, so-far performed numerically, into simpler analytic expressions. The consistency between our analytic results and the numerical integrals has been checked.

Finally, our findings simplify calculations within the Tsallis framework used in high-energy nucleus collisions and in other fields of research, where non-extensive statistics happen to be relevant.

\section*{Acknowledgments}

T.~B.~wishes to acknowledge the UCT-URC for supporting his work. S.~M.~would like to acknowledge the hospitality of the Physics Department of the University of Cape Town during his visit, as well as the indirect support from the Centre for International Teacher Education -- Cape Peninsula University of Technology.



\begin{thebibliography}{99}

\bibitem{NatureTirnakliBorges} U.~Tirnakli, and E.~P.~Borges,
Sci.\ Rep.\ {\textbf{6}} (2016) 23644.

\bibitem{GranularCombeEtAll} G.~Combe, V.~Richefeu, M.~Stasiak, and A.~P.~F.~Atman,
Phys.\ Rev.\ Lett.\ {\textbf{115}} (2015) 238301.

\bibitem{ExpeNonHEPConfirmationDouglasEtAll} P.~Douglas, S.~Bergamini, and F.~Renzoni,
Phys.\ Rev.\ Lett.\ {\textbf{96}} (2006) 110601.

\bibitem{TsallisOriginal} C.~Tsallis,
J.\ Statist.\ Phys.\ {\textbf{52}} (1988) 479.

\bibitem{TheorPredicConfirmedLutz} E.~Lutz,
Phys.\ Rev.\ A {\textbf{67}} (2003) 051402.

\bibitem{STAR} B.~I.~Abelev et al. (STAR collaboration),
Phys.\ Rev.\ C {\textbf{75}} (2007) 064901.

\bibitem{PHENIX1} A.~Adare et al. (PHENIX collaboration),
Phys.\ Rev.\ D {\textbf{83}} (2011) 052004.

\bibitem{PHENIX2} A.~Adare et al. (PHENIX collaboration),
Phys.\ Rev.\ C {\textbf{83}} (2011) 064903.

\bibitem{ALICE_charged} K.~Aamodt, et al. (ALICE collaboration),
Phys.\ Lett.\ B {\textbf{693}} (2010) 53.

\bibitem{ALICE_piplus} K.~Aamodt, et al. (ALICE collaboration),
Eur.\ Phys.\ J.\ C {\textbf{71}} (2011) 1655. 

\bibitem{CMS1} V.~Khachatryan, et al. (CMS collaboration),
JHEP {\textbf{02}} (2010) 041.

\bibitem{CMS2} V.~Khachatryan, et al. (CMS collaboration),
Phys.\ Rev.\ Lett.\ {\textbf{105}} (2010) 022002.

\bibitem{ATLAS} G.~Aad, et al. (ATLAS collaboration),
New J.\ Phys.\ {\textbf{13}} (2011) 053033.

\bibitem{ALICE_PbPb} B.~Abelev, et al. (ALICE collaboration),
Phys.\ Rev.\ Lett.\ {\textbf{109}} (2012) 252301.

\bibitem{E+E-} I.~Bediaga, E.~M.~F.~Curado, J.~M.~de~Miranda,
Physica A {\textbf{286}} (2000) 156.

\bibitem{R1} G.~Wilk and Z.~W\l{}odarczyk,
Acta Phys.\ Polon.\ B {\textbf{46}} (2015) 1103.

\bibitem{R2} K.~\"Urm\"ossy, G.~G.~Barnaf\"{o}ldi, T.~S.~Bir\'{o},
Phys.\ Lett.\ B {\textbf{701}} (2011) 111.

\bibitem{R3} K.~\"Urm\"ossy, G.~G.~Barnaf\"{o}ldi, T.~S.~Bir\'{o},
Phys.\ Lett.\ B {\textbf{718}} (2012) 125.

\bibitem{Ijmpa} P.~K.~Khandai, P.~Sett, P.~Shukla, V.~Singh,
Int.\ Jour.\ Mod.\ Phys.\ A {\textbf{28}} (2013) 1350066.

\bibitem{Plbwilk} B.~-C.~Li, Y.~-Z.~Wang and F.~-H.~Liu,
Phys.\ Lett.\ B {\textbf{725}} (2013) 352.

\bibitem{Marques} L.~Marques, J.~Cleymans and A.~Deppman,
Phys.\ Rev.\ D {\textbf{91}} (2015) 054025.

\bibitem{WilkWlodarczyk} G.~Wilk and Z.~W\l{}odarczyk,
Phys.\ Rev.\ Lett.\ {\textbf{84}} (2000) 2770.

\bibitem{Dauxois} T.~Dauxois,
J.\ Stat.\ Mech.\ (2007) N08001.

\bibitem{RapisardaPluchinoToDauxois} A.~Rapisarda and A.~Pluchino,
Eur.\ Phys.\ News {\textbf{37/2}} (2006) 10.

\bibitem{TsallisToDauxois} C.~Tsallis,
arXiv:0712.4165.

\bibitem{HilhorstSchehr} H.~J.~Hilhorst and G.~Schehr,
J.\ Stat.\ Mech.\ (2007) P06003.

\bibitem{WilkOsada} T.~Osada G.~Wilk,
Phys.\ Rev.\ C {\textbf{77}} (2009) 044903.

\bibitem{TsallisIndore} T.~Bhattacharyya, J.~Cleymans, A.~Khuntia, P.~Pareek and R.~Sahoo,
Eur.\ Phys.\ J.\ A {\textbf{52}} (2016) no.\ 2, 30.

\bibitem{Alberico} W.~M.~Alberico, A.~Lavagno, and P.~Quarati,
Nucl.\ Phys.\ A {\textbf{680}} (2001) 94c.

\bibitem{Smirnov} V.~A.~Smirnov,
{\it{Evaluating Feynman Integrals}}, Springer-Verlag Berlin Heidelberg, Germany 2004.

\bibitem{BoosDavydychev} E.~E.~Boos and A.~I.~Davydychev,
Teor.\ Mat.\ Fiz.\ {\textbf{89}} (1991) 56 [Theor.\ Math.\ Phys.\ {\textbf{89}} (1991) 1052].

\bibitem{DavydychevTausk} A.~I.~Davydychev and J.~B.~Tausk,
Nucl.\ Phys.\ B {\textbf{397}} (1993) 123.

\bibitem{Worku1} J.~Cleymans and D.~Worku,
J.\ Phys.\ G {\textbf{39}} (2012) 025006.

\bibitem{Worku2} J.~Cleymans and D.~Worku,
Eur.\ Phys.\ Jour.\ A {\textbf{48}} (2012) 160.

\bibitem{Worku3} J.~Cleymans, G.~I.~Lykasov, A.~S.~Parvan, A.~S.~Sorin, D.~Worku,
Phys.\ Lett.\ B {\textbf{723}} (2013) 351.

\bibitem{Azmi1} M.~D.~Azmi and J.~Cleymans,
J.\ Phys.\ G {\textbf{41}} (2014) 065001.

\bibitem{Ishihara} M.~Ishihara,
Int.\ Jour.\ Mod.\ Phys.\ E {\textbf{24}} (2015) no. 11, 1550085.

\bibitem{bateman} H.~Bateman,
{\it Higher Transcendental Functions}, McGraw-Hill Book Company, New York 1953 (See Section 2.10 formula (4) and the conditions therein)

\end{thebibliography}
\end{document}